\documentclass[runningheads,a4paper]{llncs}

\usepackage{amssymb}
\usepackage{graphicx}
\usepackage{multirow}
\usepackage{color,soul}
\usepackage{algorithm}
\usepackage{algorithmic}
\usepackage[]{amsmath}
\usepackage[]{epsfig}
\usepackage{enumerate}
\usepackage{color}
\usepackage{wrapfig}
\usepackage{subfig}
\usepackage{esint}
\usepackage{epstopdf}
\usepackage{epsfig}
\usepackage{qcircuit}

\usepackage{url}
\urldef{\mailsa}\path|{alfred.hofmann, ursula.barth, ingrid.haas, frank.holzwarth,|
\urldef{\mailsb}\path|anna.kramer, leonie.kunz, christine.reiss, nicole.sator,|
\urldef{\mailsc}\path|erika.siebert-cole, peter.strasser, lncs}@springer.com|    
\newcommand{\keywords}[1]{\par\addvspace\baselineskip
\noindent\keywordname\enspace\ignorespaces#1}

\begin{document}

\mainmatter  

\title{Lecture Notes in Computer Science:\\Authors' Instructions
for the Preparation\\of Camera-Ready
Contributions\\to LNCS/LNAI/LNBI Proceedings}

\titlerunning{Lecture Notes in Computer Science: Authors' Instructions}

%
%

\title{Building a Completely Reversible Computer}

\author{Martin~Lukac$^1$
\and Gerhard W. Dueck$^2$
\and Michitaka~Kameyama$^3$
\and Anirban~Pathak$^4$
\institute{$^1$Department of Computer Science, Nazrabayev University, Astana, Kazakhstan, Email:martin.lukac@nu.edu.kz\\
$^2$Faculty of Computer Science, University of New Brunswick, Fredericton, Canada, Email: gdueck@unb.ca\\
$^3$Department of Informatics, Ishinomaki Senshu University, Ishinomaki, Japan, Email: michikameyama@isenshu-u.ac.jp\\
$^4$Department of Physics and Materials Science Engineering, Jaypee Institute of Information Technology, A-10, Sector 62, UP 201307, India, Email: anirban.pathak@jiit.ac.in}
}

%
\authorrunning{Building a Completely Reversible Computer}


%
%

\toctitle{Building a Completely Reversible Computer}
\tocauthor{Lukac M., Dueck G., Kameyama M., Pathak A.}
\maketitle

\begin{abstract}
A critical analysis of the feasibility of reversible computing is performed. The key question is: Is it possible to build a completely reversible computer? A closer look into the internal aspects of the reversible computing as well as the external constraints such as the second law of thermodynamics has demonstrated that several difficulties would have to be solved before reversible computer is being built. 
It is shown that a conventional reversible computer would require energy for setting up the reversible inputs from irreversible signals, for the reading out of the reversible outputs, for the transport of the information between logic elements and finally for the control signals that will require more energy dissipating into the environment. A loose bound on the minimum amount of  energy required to be  dissipated during the physical implementation of  a reversible computer is obtained and a generalization of the principles for reversible computing is provided. 
\end{abstract}

\section{Introduction}
\label{sec:1}

In the last two decades, a considerable effort has been directed to the study of reversible computers as one of the possible alternatives to the currently most popular CMOS-based irreversible computing technology. 
Landauer~\cite{landauer:61} showed that erasure of any bit results in the dissipation of energy. 
Additionally, the interconnect and wires in logic circuits  also dissipate heat due to resistive dissipation which is increasing not only with higher VLSI but also with higher processing speeds. 
This problem is moreover accentuated by Moore's law~\cite{moore:65} that predicts the doubling of the number of transistors per square unit area in every 18th month. 

The combined effect of these heat dissipation mechanisms poses a serious challenge to the current level of VLSI.
Any further development would require new architectures. 
Innovative technologies are to be sought in order to allow further development of computing devices 
with increased performance.

A low power and low heat dissipation replacement of the current CMOS based computing devices has been actively sought. In this context, several separate features of reversible computing have been shown to be superior to classical computing model. For instance information reversibility has been effective for lost state recovery in sequential logic~\cite{lukac:12}, testing reversible circuits is much simpler~\cite{wille:13}, adiabatic charging~\cite{haenninen:13}) reduces power of computing.
Despite the fact that, in theory, reversible computing is a concept that 
offers solutions at various levels of implementation, it has not captured the mainstream research and development. 
. 
The dissipation is 
the result of the loss of energy due to switching as well as (and mostly) due to the $I^2R$ transmission heat resistive dissipation that happens through the large number of interconnections present in ICs. 
Reversible computing offers some considerable features of computing and power savings but also has several restrictions in implementation. Considering the heat dissipation and the potential feats of reversible computing, it is reasonable to ask: can reversible computing exists in a irreversible world or how much a reversible computer can be reversible in theory and in practice?

In this paper, we examine the concept of reversible computing from a realization point of view. We aim to study  various computational stages required for reversible computing and check whether all these stages are completely reversible.
 Moreover, we would like to take a closer look at the energy dissipation and information loss  that may exist in the possible realizations of reversible circuits.
Theoretical and practical constraints standing in the way of realizing reversible computers are also considered. 
Finally, we look at the different stages of a reversible computer and provide a rigorous description of energy dissipation. 

This paper is organized as follows. Section~\ref{sec:back} introduces concepts related to reversible functions and computers. Section~\ref{sec:edissip} discusses the concepts of energy dissipation in computing systems and Section~\ref{sec:closed} discusses the so called closed computational systems. Section~\ref{sec:transf} analyses the energy transfer computational systems. Section~\ref{sec:qcom} discusses the possibilities of building a reversible computer in quantum computing technology and Section~\ref{sec:steps} looks in more details at the individual steps of dissipation in reversible computers. Finally Section~\ref{sec:con} concludes this papers. 

\section{Background}\label{sec:back}
\begin{definition}[Reversible Logic Function] 
	A function $f:A\rightarrow B$ is reversible if it is bijective (i.e., one-to-one and onto) and thus $f^{-1}:B\rightarrow A$. In other words, $f(X_a) = f(X_b) \implies X_a = X_b$.

\end{definition}

Table~\ref{tab:revunrev} shows examples of reversible and irreversible functions.
\begin{table}[bht]
\centering
\caption{\label{tab:revunrev} Example of (a) reversible logic function (CNOT) and (b) irreversible logic function.}
\begin{tabular}{ccc}
\begin{tabular}{|c|c|}
\hline
AB&A'B'\\
\hline
00&00\\
01&01\\
10&11\\
11&10\\
\hline
\end{tabular}
& &
\begin{tabular}{|c|c|}
\hline
AB&A'B'\\
\hline
00&01\\
01&11\\
11&00\\
10&11\\
\hline
\end{tabular}\\
(a)& &(b)
\end{tabular}
\end{table}



\begin{definition}[Conservative Reversible Function]
	A function $f:I\rightarrow O$ is said to be conservative reversible if it is bijective ($f^{-1}:O\rightarrow I$) and if for every tuple $(I_i,O_i)$, $\vert I_i(k)\vert = \vert O_i(k)\vert$ for $k=0,\ldots,r-1$, with $I(O)_i$ being the $i^{th}$ input (output) to (from) the function and $\vert I(O)_i(k)\vert$ is the number of bits with value $k$ in the encoding term. 
\end{definition}

Table~\ref{tab:consrev} shows an example of a conservative reversible function.
\begin{table}
\centering
\caption{\label{tab:consrev} Example of (a) conservative reversible logic function and (b) non-conservative reversible logic function}
\begin{tabular}{ccc}
\begin{tabular}{|c|c|}
\hline
AB&A'B'\\
\hline
00&00\\
01&10\\
10&01\\
11&11\\
\hline
\end{tabular}
& &
\begin{tabular}{|c|c|}
\hline
AB&A'B'\\
\hline
00&11\\
01&10\\
10&00\\
11&01\\
\hline
\end{tabular}\\
(a)& &(b)
\end{tabular}
\end{table}

To build an arbitrary reversible function in reversible circuit a set of universal reversible gates is required.
\begin{definition}[Universal gate set]
is a set of gates that allows to build arbitrary Turing-computable function.
\end{definition}

\begin{figure}
	\centering
	\includegraphics[width=\linewidth]{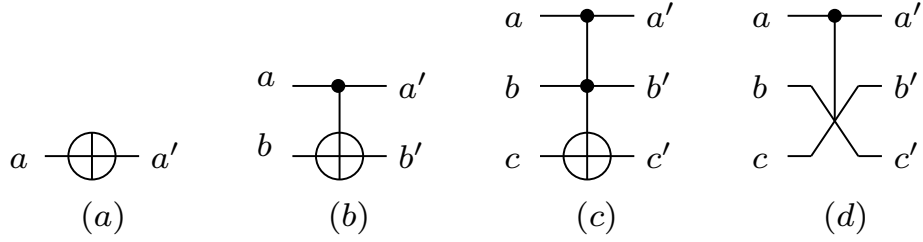}
	\caption{\label{fig:usets} Different reversible gates (a) $NOT$, (b) $CNOT$, (c) $C^2NOT$, (d) $CSWAP$}
\end{figure}

The simplest classical reversible universal gate set includes the Toffoli (also known  as $C^2NOT$) gate and negation ($NOT$ gate) or $C^2NOT$ gate and constant inputs. Both gates are shown in Figure~\ref{fig:usets}(a) and (c). 
Another universal gate set contains the $Fredkin$ gate (also known as $CSWAP$) and the $NOT$ gate, shown in Figure~\ref{fig:usets}(a) and (d). Unlike the $C^2NOT$ gate the $CSWAP$ gate is also conservative. 
CSWAP gates have been used for building of conservative reversible circuits~\cite{fredkin:82,maslov:05,donald:08}. 

Logical reversibility is in general not a strong enough motivation to build reversible computers. Significant benefits can be obtained if the devices are naturally logically reversible or energy recycling~\cite{teichmann:11,arfin:12} are used to build reversible computational devices. 
Evolution of states is inherently reversible in quantum mechanics. Computing performed using quantum resources are referred to as quantum computer which has received significant attention in the last decade. A quantum computer uses quantum gates  similar to their classical counter parts. A few of such gates are  shown in Eq.~\ref{eq:qgates}.
\begin{equation}
\begin{split}
R_x(\theta)=\begin{bmatrix}\cos(\frac{\theta}{2})& -i \sin(\frac{\theta}{2})\\-i \sin(\frac{\theta}{2})& \cos(\frac{\theta}{2})\end{bmatrix},& H=\frac{1}{\sqrt{2}}\begin{bmatrix}1&1\\1&-1\end{bmatrix}\\
I_{ZZ}(\theta) = e^{i\frac{\theta}{2}}\begin{bmatrix}e^{-i\frac{\theta}{2}}&0&0&0\\0&e^{i\frac{\theta}{2}}&0&0\\0&0&e^{i\frac{\theta}{2}}&0\\0&0&0&e^{-i\frac{\theta}{2}}\end{bmatrix},& T=\begin{bmatrix}1&0\\0&e^{i\frac{\pi}{2}}\end{bmatrix}
\end{split}
\label{eq:qgates}
\end{equation}
In this paper we will refer to three main types of technologies. The irreversible CMOS and transistor based technology, dual-lines transistor and switch based technology (such as the one proposed in \cite{desoete:02}) and finally quantum technology. 
Each technology will be used to illustrate some particular properties of 
reversible systems.

\begin{definition}[Closed reversible computation (CRC)]
is a reversible computational system that is executed within a reversible environment.
\label{def:crc}
\end{definition}

\begin{definition}[Reversible transfer computation (RTC)]
is a reversible system that is executed from within an irreversible environment.
\label{def:rtc}
\end{definition}

The main difference between the systems defined in def~\ref{def:crc} and~\ref{def:rtc} is that the first system does not dissipate any energy and cannot be directly operated. This is assuming that interacting with the reversible system requires energy dissipation. The system from def.~\ref{def:rtc} is such system that allows toinitiate and observe results from within an irreversible environnet while the computation is isolated and does not dissipate energy.

\begin{definition}[Reversible Inputs]
is a transformation required to obtain signals acceptable by a reversible circuit.
\end{definition}

For instance, in case of quantum computation, classical signals have to be transformed into quantum states. In dual rail CMOS technology the signal has to be complemented. Naturally, if the reversible system uses normal electrical signals, the reversible signal is only a label for the input to the reversible circuit.

\begin{definition}[Energy reversible computation]
is a computational process that either:
\begin{itemize}
	\item allows to recover some amount energy $\Delta E$ used while computing $f$ by computing $f^{-1}$;
	\item allows to adiabatically recover energy 
		$\Delta E$ used to calculate $f$;
	\item allows to recover energy $\Delta E$ used to calculate $f$ by capturing the dissipated energy and the used energy back to the power supply.
\end{itemize}
\end{definition}

A reversible computer now can be defined over multiple levels (classes) of reversibility.
\begin{enumerate}
    \item Not sustained logical reversibility (NSLR): reversibility allowing to recover inputs from outputs without any device requirement. The NSLR requires a software components to track all element of the logic circuits required to recover original input from the output values. 
    \item Sustained logical reversibility (SLR): reversibility resulting from the reversibility of hardware components. This level represents a computr that is build from components that are logically reversible but the computer is not intended to be logically reversible. 
    \item Energy sustained reversibility (ESR): reversibility from energy reversible or conservative hardware components. This level represent a compter that is built from components tht recycle energy and thus are energy cpnservative. 
    \item Fully sustained reversibility (FSR): reversibility resulting from energy reversible logical components and ideal transmission elements. 
%
\end{enumerate}

In addition to the reversibility of the computing system, we will also consider the environment. 
In particular, we look closer to what kind of interaction exists between the reversible system and the operator (the environment).

For most the rest of the paper, we will consider the environment as an open system; i.e. energy dissipated through either the switching or the thermal dissipation process in general cannot be captured but is rather diffused into the environment. 

Finally, we will consider as the baseline of computers and computation the current state of art, transistor based CMOS technology. 
We assume that the normal input being represented by a voltage high (low) and output as well. 

\section{Energy Dissipation in Generalized Computing Processes}
\label{sec:edissip}

\begin{figure}
	\centering
	\includegraphics[width=0.6\linewidth]{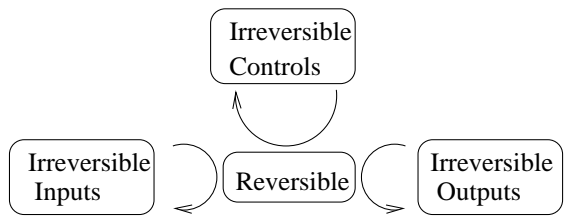}
	\caption{\label{fig:revers} The interaction between the reversible computer and irreversible world}
\end{figure}

It has be established how a reversible computer co-exists in the irreversible world.
By default an information processing system requires an input and an output. 
Depending on the level of reversibility of the computing system the input setting and output reading  have different requirements.  
In general, a reversible computer must interact with several irreversible sources of information. 
This is shown in Figure~\ref{fig:revers} where several irreversible components are shown interacting with a reversible computation.

Setting an input in a reversible system means adding or removing energy. 
Consider a CRC with equivalent internal states, i.e., each possible configuration of inputs, computation and outputs have the same energy. 
Then setting inputs and control signals directly changes the output values by simply reconfiguring a set of the input values to the output with some configured function. 


There are two possible interaction with a fully sustained reversible computational system:
\begin{enumerate}
	\item The reversible system is closed and is within a closed environment.
	\item The reversible system directly interacts with irreversible world and the interaction is energy dissipative.
\end{enumerate}

For case 1): A fully sustained reversible computation is conservative (by definition). If it is conservative it implies that it must be closed otherwise the energy of the computational system would be changing. 
For case 2): A fully sustained reversible computational is conservative and contains all possible combinations of inputs-to-outputs mappings. 
Then reconfiguring the system from one input to another requires energy dissipation or adding energy  in order to provide the work required for changing the system's configuration.

%
%



\section{Closed Reversible System}
\label{sec:closed}

In an ideal case, a reversible system could be viewed as a  closed system because a truly reversible system would conserve energy and consequently,  it must be restricted from transferring energy to   the environment. In contrast, an open system allows interaction of the system with the environment which is usually modeled as a bath \cite{breuer:02}. In reality, it is impossible to completely isolate a system from the environment. Thus, to some extent every system is open system, and consequently, the system-bath interaction would lead to some amount of dissipation of energy. However, for the computational model building, we may assume for a while that a closed system exists (or a system exists for which system-bath coupling is negligibly small), which does not allow energy of system to dissipate into environment. In such a situation, a closed system can be viewed as a system consisting of  $r^n$ computing configurations; each configuration is represented by a particular energy conservative computing configuration mapping combination of inputs to outputs and an energy dissipating process allowing to switch between these configurations.  
\begin{figure}[bht]
	\centering
	\includegraphics[width=0.6\linewidth]{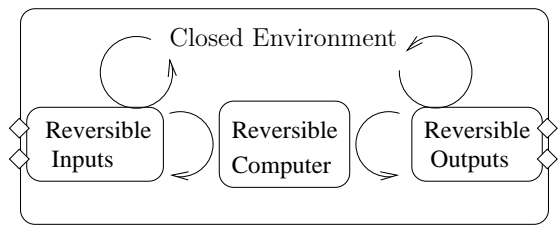}
	\caption{\label{fig:closed} Schematic representation of the closed world model of reversible computer}
\end{figure}

In fact, an open system can also be viewed as a closed system, but the boundary of environment is at a large distance from the system (in other words, dimension of the environment is much larger than that of the system).  Actually, in context of open system, specially, in quantum cases, we often use two approximations-  the Born approximation and the Markov approximation \cite{breuer:02}. Born approximation assumes a weak coupling between the system and the bath, it also assumes that the bath is very large and therefore, it is not affected by the system. This approximation is very much realistic.  Further, the assumption that the bath does not have a memory is known as Markov approximation. System bath-interaction and loss of energy and/or information are frequently studied under both Markovian and non-Markovian approximation. It is not our purpose to discuss them in detail. We just wish note that in reality, it is impossible to construct a closed system (a closed reversible computer) and thus to circumvent loss of energy/information.




The CRC must naturally be conservative and ideal in the sense that the change of configuration of the inputs and outputs represent the all required energy for the system. 

\section{Reversible Transfer Computation}
\label{sec:transf}

 Basic idea of open system has already been given in the previous section. In the open system world, the dictating principle of the 2nd law of thermodynamics does not allow to reverse most of the actions. In a general computational process that includes all the required steps from input and computer initiation to the output read-out the dissipation happens differently at almost every step. 

A RTC is such a system that accepts irreversible inputs, but its operation is reversible. The general scheme is shown in Figure~\ref{fig:revers}. An example of such system is the switch based technology~\cite{devos:94,vanrentergem:05}. Such system is similar in essence to the closed systems in principle, but is intended to work directly with irreversible information. 

\begin{figure}
	\centering
	\includegraphics[width=0.6\linewidth]{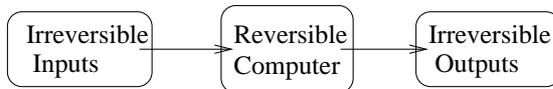}
	\caption{\label{fig:direct} Schematic representation of a direct interaction process with a reversible computer}
\end{figure}


\section{Quantum Computing}
\label{sec:qcom}

In non-relativistic quantum mechanics, a state follows Schrodinger equation which in turn (along with the fact that outcomes of a measurement has to be real) implies that the evolution of the state is described by a unitary operator. Unitary evolution automatically implies reversibility as existence of a unitary operator $U$ implies the existence of $U^{\dagger}=U^{-1}$.  As a consequence quantum mechanical operations are intrinsically reversible in nature and all quantum computers are reversible. There exist various proposals for building quantum computer \cite{nielsen:00}, and perhaps quantum computer is the most interesting and most discussed reversible computer. Although, evolution of states are reversible here, the measurement operators and the Kraus operators that describes affects of different Markovian \cite{thapliyal:15} and non-Markovian \cite{thapliyal:16} noise models are not unitary. In any realistic situation,  decoherence or noise will be present and that  would ensure that a quantum computer would not operate as a completely reversible manner. Further, some implementations of gates (see KLM approach \cite{knill:01}), measurement of the outcome of computation and error correction \cite{pathak:13} would require measurement (i.e., applications of non-unitary or irreversible operations) either on the qubits involved in computation or on the ancilla qubits and prohibit us from constructing a completely reversible quantum computer. Basis difference between a classical reversible and a quantum computer lies in the fact that the gates involved in quantum computation can accept superposition states as the input (specially, an input superposition state in a controlled qubit (say, $\alpha|0\rangle+\beta|1\rangle$ in the  controlled qubit of a $CNOT$ gate) would lead to an entangled state having no classical analogue). 

\begin{figure}
	\centering
	\includegraphics[width=0.5\linewidth]{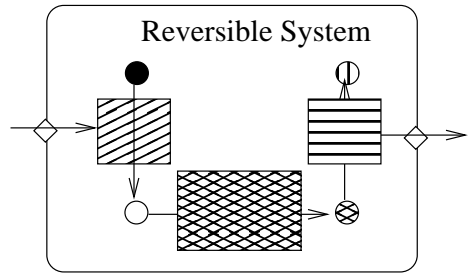}
	\caption{\label{fig:qinmeas} Schematic representation of the initiation, computing and measurement executed in the framework of a closed quantum environment}
\end{figure}

In fact, in absence of noise, a quantum computer approximates very closely the schematic representation shown in Figure~\ref{fig:closed}. Further, and most importantly, in analogy with its classical counter part, the reading out of the outputs and error correction would always involve some dissipation of energy and a schematic representation of this process is shown in Figure~\ref{fig:qinmeas}. In brief, if we assume that we can isolate a quantum computer from the environment  and avoid all types of decoherence (which is practically impossible as decoherence is the major problem in the experimental realization of a scalable quantum computer), still it will not be completely reversible as the input setting (in most cases) and outputs reading (always) would involve the application of nonunitary (equivalently irreversible) operations (Section~\ref{sec:inout}). 


\section{Steps of Dissipation}
\label{sec:steps}

In a computing system, several steps lead to and result in  dissipation. 
In reversible computers various dissipation processes occur. 
We look closely at four different processes that are in general required to interact with a reversible computer: 
\begin{enumerate}
\item  irreversible inputs to reversible inputs;
\item  reversible outputs to irreversible outputs;
\item  irreversible controls to reversible instructions;
\item  reversible computations.
\end{enumerate}

 \subsection{Input Initialization and Output Retrieval}
 \label{sec:inout}
 A reversible computer requires a reversible input. 
 Depending on the implementing technology the input realization will have different specifications~\cite{nielsen:00}. 
 In quantum implementations the classical input must be transformed into specifically prepared elementary particles representing the input state~\cite{lukac:12}. In a switched based approach such as~\cite{devos:95} the irreversible input must be transformed into a dual rail signal. 

\begin{figure}
	\centering
	\includegraphics[width=0.5\linewidth]{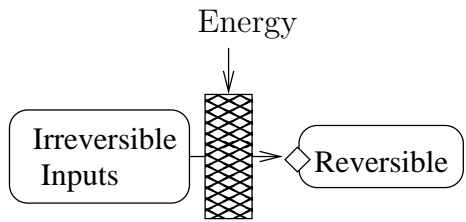}
	\caption{\label{fig:setinputs} The process of transforming irreversible inputs to reversible input values is a naturally dissipative process.}
\end{figure}

The irreversible inputs read operation dissipates energy according to Landauer's principle; $kT\,ln(2)$ and every bit of transcription dissipates such amount of energy. 
This information dissipation occurs always at some level of interaction between irreversible and reversible computer. 
The input setting process is schematically depicted in Figure~\ref{fig:setinputs}. 

Output reading requires transformation of $0$ to $1$ or $1$ to $0$ that increase or decrease of energy respectively. Assuming that the output information is always 0, the reading of the output requires energy from the environment in order to generate correct irreversible outputs. 

Similarly to the input, the output must also be in most of cases of reversible hardware extracted from the reversible computer. This transformation based reading is most remarkable again in quantum computing where the output must be measured. The measurement of a quantum system destroys the quantum information and preserves only the values projected on observable axes. Moreover, the measurement is not a unitary transformation and thus breaks the general reversibility of the whole quantum system. Consequently, both the initialization and the measurement in a quantum computer both are dissipative processes. 

\subsection{Control Signals}
The reconfiguration of the circuit implementing the logic also enters into into consideration when the input variables are used to control logic switches. If the switch change requires energy then it either dissipates for each configuration or each switch is connected to a reversible circuit. 
Thus similarly to the input setting the dissipation occurs when using irreversible information to set the reversible control signals. 
\begin{figure}
	\centering
	\includegraphics[width=0.4\linewidth]{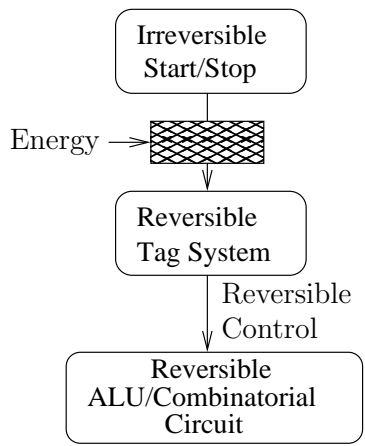}
	\caption{\label{fig:revcontrol} Example of reversible instructions implemented as a cyclic tag system controlled by a single irreversible variable.}
\end{figure}
However, one could assume that the control logic is implemented by a cyclic tag system~\cite{cook:04} simulated by a reversible Turing machine; such machine is only initialized externally and then sequentially executes a set of instructions. Such system can be seen as a reversible instruction set implemented in a reversible circuit as schematically shown in Figure~\ref{fig:revcontrol}.

\subsection{Computing}
\label{sec:com}
Thus far we have focused on analyzing the interactions of the reversible computer with the environment. In this section, we look in detail at various dissipation components of the reversible computer itself. 

\begin{figure}
	\centering
	\includegraphics[width=0.6\linewidth]{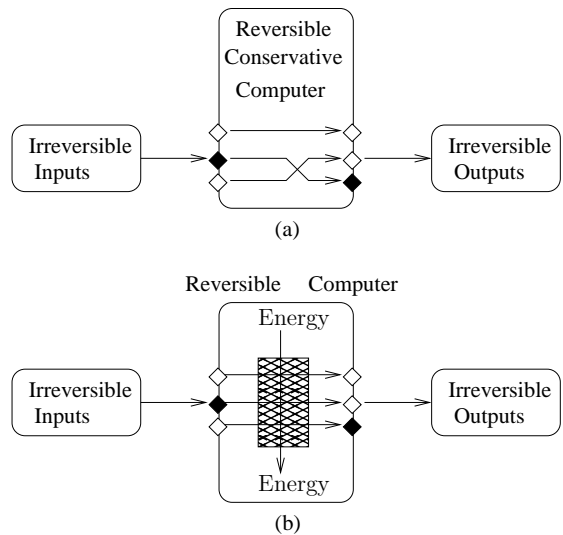}
	\caption{\label{fig:computer} Schematic representation of (a) Conservative Reversible Computer  and (b) Non conservative Reversible computer with internal state changes}
\end{figure}

During computation and depending on the level of reversibility, the computational system uses the input-output reversible mapping which is either energy conservative or non conservative process. 
In the case of being a conservative process, the input-to-output mapping can be performed by the permutation of input signals and thus the only dissipation of the reversible computer is given by the configuration of the permutation of control signals (Figure~\ref{fig:computer}(a)). 
If the reversible computer is not conservative the output signals cannot be obtained by a permutation of input signals and thus states of individual input variable have to be changed. Such changes however could be considered as being internal to the system and thus be reversible if coupled in a closed system as shown in Section~\ref{sec:closed} (Figure~\ref{fig:computer}(b)).

Note that every reversible computer can be implemented with at most $2n$ input bits as a conservative computer. This is simple to see: a reversible computer can at most transform its input to their negation. Consequently, having $n$ input bits and $n$ input complement bits allows to realize any reversible function by implementing $F:f_r\circ f_{\bar{r}}$. For instance assume that one wants to implement a Toffoli function. 

The interconnection adds significantly to the energy/infor\-mation dissipation problem.
Whether it concerns the wires between logic components, high speed buses, optical beams or state transfers, these information displacements affect the amount of dissipated energy and information. In a classical wire based implementation of a reversible computer the $KT\,ln(2)$  energy dissipation is much lower than the $I^2R$ thermal dissipation due to the wire transport. In fact, one can estimate the scale of integration, the density of transistors and the interconnect  worth implementing as reversible computation. 

The standard power dissipation in a CPU is given in Eq.~\ref{eq:pow}.
\begin{equation}
    P_{cpu} = P_{dyn}+P_{sc}+P_{leak}.
    \label{eq:pow}
\end{equation}
with $P_{dyn}$, $P_{sc}$ and $P_{leak}$ being the dynamic, the short-circuit and the leakage power consumption~\cite{pedram:02,piguet:06,arora:10}.  Let us deal with an ideal transistor that has a dissipation specified only by the Landuaer's law, i.e. each transistor dissipates at maximum $KT\, ln(2)$ joules per bit erasure. That is the $P_{sc}+P_{leak} = 0$ and the $P_{dyn} = CV^2f$ with $CV$ being symbolically set to 1. Then in a device using ideal transistors and interconnect we can estimate how much amount of interconnect should be reduced so that the transistor dissipation is equal or larger to the thermal wire dissipation. 

The wire thermal dissipation is given by $I^2R$ where $I$ is the power and $R$ is the resistance of the wire. 
Let assume that we are dealing with a $0.12nm$ technology. 
That is let $\lambda=12$ and the space required between two adjacent transistors is set to $2\lambda$. 
Using these numbers we can estimate the conductor resistance (in this case copper) $R_{dc}=\frac{\rho\times l}{s} = \frac{1.678e^{-8}\times 2.4e^{-5}}{1.2e^{-5}}$. 
Let $T=293.15K$, $k=1.38e^{-23} J/K$ and $ln 2 = 0.693$. 
Finally let $I$ be only proportional to the clock speed by $P_{switch} = 0.5CV^2f$ with $C\approx 30fF/\mu m$, $V=5V$ and $f$ to be a parameter, we can equate the switching power with Landauer's  dissipation energy:
\begin{equation}
\begin{split}
    kT\, ln2&\leq I^2R\\
    kT\, ln2&\leq \frac{\rho l}{s}\\
    kT\, ln2&\leq \frac{\rho l}{s}\\
\end{split}
\end{equation}

\subsection{Generalization}

There are four steps where energy must be spent without any exception: setting the inputs, reading the outputs, the interconnect  and controlling the reversible computer.

The first and second cases are the setting and reading  of the irreversible inputs and reversible outputs respectively to and from the reversible circuit. 
This case is not avoidable because the interfacing with a reversible circuit will at certain point in the sequence input setting-computation-output reading has to dissipate energy to transform irreversible signals to reversible signals or vice versa.

The energy dissipation in these two steps can have either only an electrical component (such as encoding and decoding) or can also require a physical component such as additional signal wires. 

The problem of interconnect has been addressed in Section~\ref{sec:com} and so far there is no such technology that would allow a cost free information transfer between logic gates. 

The fourth case is when the reversible computer must be controlled in anyway by an irreversible clock or control signals. Again, this is in general necessary in order to synchronize the computer with specific input time schedule. This is required for instance if the input is a real time sequence of inputs from sensors or dynamically changing environment. 

Finally, it is important to notice that given the previously described requirements the advantages of a reversible computer and possible savings can be estimated. Thus, let $k$ be the number of input bits, $l$ number of output bits, $i_r$ be the instruction register, and $p_r$ a reversible combinatorial circuit. Then in order to implement a irreversible function, the energy loss at least $2nKTln(2)$ with $2n = k+l+i_r+n_{p_r}$ in the simplest case, with $n_{p_r}$ being the number of qubits in the combinatorial circuit $p_r$. This value represents a lower bound on the constraints of a reversible computer presented in this paper. As it can be seen the $2nKTln(2)$ is the amount of energy lost if the reversible computer is implemented as a self-contained cyclic reversible tag system. The $KTln(2)$ is the Landauer single irreversible bit operation energy dissipation and $2n$ is the minimum number of irreversible single bit operations required to initialize the inputs and reading the outputs. Due to the components of reversible computer taken into account each with $k$, $l$, $i_r$ and $p_r$ qubits the simplest possible logic operation is the flip of all qubits resulting in $2n$ changes of states of the $2n$ qubits.

\section{Conclusion}
\label{sec:con}

In this paper, we have discussed the possibility of building a completely reversible computer in the real world. In particular, we have established that all the tasks performed in a computing machine cannot be performed reversibly, and thus a computer cannot be completely reversible. In other words, it would always involve some energy consumption. This is expected as it is in consistency with the second law of thermodynamics. What is interesting here is the point that specific tasks (or parts of the computing tasks) that cannot be performed in a reversible manner are identified. Further, the minimum energy required for performing those tasks has been computed to yield a lower bound on the energy dissipation that happens in a reversible computer embedded in an irreversible environment.  In addition, it may be noted that even if we succeed in future to synthesize meta-materials that can show superconductivity at room temperature, and replace all the interconnecting wires present in a computer/ICs by wires made of this meta-material to circumvent $I^{2}R$ loss, we will not have a completely reversible computer as the other sources of power dissipation mentioned here would still remain present. Thus, even if a reversible computer is built with superconducting interconnections or lossless fibers (in case of optical implementations), it will not be a perpetual machine. Rather, it will honor second law of thermodynamics and dissipate energy through certain steps listed in the previous sections. Only difference between such a reversible computing machine and a conventional irreversible computing machine would be that the reversible one would require lesser amount of energy. The amount of energy that can be saved by doing computation in reversible manner may appear negligible at the moment when $I^{2}R$ loss is significantly high, but in future if a room temperature superconducting material is discovered, then in absence of $I^{2}R$ loss, the energy saving through reversible operation would appear to be significant. Further, a specific type of reversible computer, namely quantum computer (discussed in Sec. \ref{sec:qcom}) is expected to overrule all classical computers because it can solve certain computational problems (say, factorization of a large number \cite{shor:97}, and unsorted database search \cite{grover:96}) in a speed which is not achievable by its classical counter parts (for detail see \cite{pathak:13,nielsen:00}). Keeping all these facts in mind, we conclude this article by noting that it is not possible build a completely reversible classical/quantum computer, but the partially reversible machines where the main computation is done reversibly will also be of much use because of the energy saving in the classical case and computational power in the quantum case.

\vspace{2em}
\textbf{Acknowledgment:}  AP
thanks Department of Science and Technology (DST), India for the support
provided through the project number EMR/2015/000393.

\bibliographystyle{splncs03}
\small
\bibliography{../main}

\end{document}